\newcommand{\nc}{\nocite}
\newcommand{\vb}{\vspace{-7 pt}}

\newcommand{\dirac}{\!\!\!\not{\!\partial}}
\newcommand{\dir}{\!\!\!\!\!\!\not{\partial}}

\newcommand{\beq}{\begin{equation}}
\newcommand{\eeq}{\end{equation}}
\newcommand{\bea}{\begin{eqnarray}}
\newcommand{\eea}{\end{eqnarray}}

\documentstyle[12pt]{article}
\begin{document}
\input psfig.tex
\bea
 & & \nonumber
\eea 

\vspace{3 cm}

\centerline{\bf HYPERONS AS SOLITONS IN CHIRAL QUARK MODEL\footnote{
Talk at XXVIII Rencontres de Moriond, QCD and Hadronic Interactions,
Les Arcs, March 20--27, 1993}}
\vspace{0.37truein}
\centerline{\footnotesize \underline{M. PRASZA{\L}OWICZ}\footnote{
Alexander von Humboldt Fellow, on leave of absence from the 
Institute of Physics, Jagellonian
University, ul. Reymonta 4, 30-059 Krak{\'o}w. Poland},
A. BLOTZ and K. GOEKE}
\vspace{0.015truein}
\centerline{\footnotesize\it Institute for Theor. Physics II, Ruhr-University}
\baselineskip=12pt
\centerline{\footnotesize\it  4630 Bochum, Germany}

\vspace{2 cm}

\noindent ABSTRACT
  
\vspace{0.5 cm}
  
\noindent In this talk we will discuss the phenomenology of the SU(3)
Chiral Quark Model 
in which quarks interact {\em via} a self-consistent meson field,
which takes a {\em hedgehog} soliton form.
The {\em classical} part, {\sl i.e.} the energy of the soliton,
is exactly the same as in the two flavor case.
The {\em quantum} corrections are
calculated by an adiabatic rotation of the soliton resulting in a
hamiltonian analogous to the one of the Skyrmion.
A novelty is due to the mixed
terms linear in the current quark mass and in the rotational velocity,
connected to the anomalous part of the action, which
get main contribution from the {\em valence} quarks. The resulting
spectrum fits the data with a 10~\% accuracy. At the same time
the isospin splittings due to the $m_{\rm d}-m_{\rm u}$ mass difference
are reproduced within experimental errors.
Terms of the similar nature appear for some other observables like the
axial coupling $g_{\rm A}$ and it is argued that they   cure the 
disease of too small $g_{\rm A}$ inherent for all chiral models.

\newpage
\baselineskip 15pt
\section{INTRODUCTION: EFFECTIVE MODELS}

It is still an outstanding problem to calculate the low energy properties of
the hadronic spectrum. In fact evaluating 
 some quantities in the high energy regime is in a sense  much
easier than to reproduce the old data on hadronic
masses or magnetic moments. On the one hand the {\it first principle}
calculations are plagued by enormous technical difficulties and 
on the other hand the 
calculations in the non-relativistic quark model (CQM), for instance, are
theoretically unsound, although phenomenologically in many respects
quite successful.

It is therefore tempting to try another approach, which would share
some basic features with QCD and, in the same time, support the quark
model ideas. The basic features of QCD with respect to the low energy
limit are: chiral symmetry and confinement. The basic feature of the
quark model is -- not a surprise -- the very existence of quarks. We
will present here a model in which chiral symmetry  breaking generates
non-zero constituent mass $M$ for otherwise massless quarks
coupled to Goldstone bosons$^{1),2)}$,
 \nc{cqm:Blotz2,mp:iso} but we will 
almost completely ignore confinement. Not entirely though, since one
has to assume that the physical degrees of freedom are color singlets.

Let us give some motivation. 
The generating functional  of  QCD involves quark and gluon fields.
One can  imagine the following scenario$^{3)}$: \nc{chl:Ball3} first 
integrate out gluons.  
The resulting action would then describe the 
nonlinear and nonlocal many quark  interactions.  
The  next  step 
would consist in linearizing this complicated action and  expressing  it 
in terms of local  color  singlet  composite  fields  corresponding  to 
pseudo-scalar mesons coupled in a chirally  invariant  way  to 
quark fields.  
And finally we would have to integrate out quarks  to 
end up with a pion (or $\pi$--K--$\eta$) effective lagrangian
So we go  through a chain of effective actions (see 
Refs.\cite{chl:Ball3,chl:IanMog,mp:padova} 
for review and references):
\begin{equation}
 S_{\rm QCD}[q,\, A] \rightarrow S_{\rm eff}[q] \rightarrow
S_{\rm eff}[q,\, \pi] \rightarrow S_{\rm eff}[\pi], 
\label{eq:chain}
\end{equation}

It should be kept in mind that the arrows in  Eq.(\ref{eq:chain})  do 
not indicate a rigorous derivation  of  one  action  from  another  but 
rather {\em educated} guesses based mainly on  symmetry  principles  and 
some physical input. 
As $S_{\rm eff}[q,\pi]$ we will choose a semobosonized action of
the Nambu--Jona-Lasinio (NJL) model, which
is expected to follow from QCD in the instanton liquid model of the QCD
vacuum$^{6),7)}$\nc{inst:dp1,inst:dp2}. Then we have:
\beq
S_{\rm eff}[U]=-{\rm Sp} \log 
(i \dirac \, -\, m\, -\, M \; U^{\gamma_{5}}). 
\label{eq:Seff}
\eeq
where $U^{\gamma_{5}}$ describes chiral fields $\pi$ (or $\pi$-K-$\eta$)
and $m$ denotes here the current quark matrix.
Eq.(\ref{eq:Seff}) has
to be  regularized. The regularization procedure will introduce
 another implicit parameter: the cut-off $\Lambda$.
 
 From the gradient expansion of Eq.(\ref{eq:Seff}) in meson sector 
 one fixes $m$ and additionally $\Lambda=\Lambda(M)$. 
 Therefore the model is very economical: there is in fact only one
explicit parameter, namely $M$. 
More complicated choices for $S_{\rm eff}$
have been also studied, and it seems that the results for the
observables are in principle not changed$^{8)}$ \nc{mp:ga}.

\section{BARYONS AS SOLITONS}

Now we will make crucial assumptions. We will assume that
baryons can be described as {\em solitons} of the effective action
of Eq.(\ref{eq:Seff}). That means that the Goldstone fields desribed by
matrix $U$ are, in a sense, large and fulfil classical equations of motion.
To this end we choose a {\em hedgehog } Ansatz for $U$:
\begin{eqnarray}
 U_{0}\, = \,\left[ \begin{array}{cc}
 \overline{U}_{0}   &      0 \\
    0  & 1   
\end{array} \right] 
\end{eqnarray}
where $ \overline{U}_{0}=
\cos P(r) + i \vec{n}  \vec{\tau} \sin P(r) $, 
$P(0)=\pi$ and $P(\infty)=0$
Then we  rotate $U_0$ adiabatically introducing a time-dependent SU(3)
matrix $A(t)$:
$ U_{0} \rightarrow A(t)\, U_{0}\, A^{\dagger}(t),$
where
$ A^{\dagger}\; dA/dt = i/2\;  \sum_{a=1}^{8}
\lambda_{a} \Omega_{a}.$
This procedure of introducing collective coordinates and the
quantization of the system was developed in the context of the Skyrme
model and was extensively discussed in the literature$^{9)}$
\nc{rvs:bal}.

In order to calculate baryon masses one usually considers
the Euclidean correlation function for two
baryonic currents $J\;^{10)}$:\nc{inst:nucleon}
\begin{eqnarray}
\left< J(T)\;J^+(0) \right> & \approx &
\int DU\;Dq\; Dq^+ J(T)\;J^{\dagger}(0)\; e^{ i \int d^4x\,q^{\dagger}
(~\, \,\,\, \dir + m\, +\, M \; U^{\gamma_{5}})q} \nonumber\\
 & \approx & \Gamma^f \Gamma^g \int DU 
 \prod\limits_{i=1}^{N_{\rm c}}G_{U}^{f_i,g_i}(T,0)\, 
 e^{-S_{\rm eff}[U]}  \nonumber \\
& \approx & e^{-T(E_{\rm level}+E_{\rm field})}~~~~~~
 \approx~~~~~  e^{-T M_{\rm Baryon}} \label{eq:JJ}
\end{eqnarray}
where $\Gamma$ denotes schematically the projection of $N_{\rm c}$
quarks on the color singlet baryonic   state and $G_U$ is the quark
propagator in the presence of the mean field $U$. 
From Eq.(\ref{eq:JJ}) one has to subtract the {\em reference} energy 
of the vacuum (no soliton), and apply the regularization procedure.
Equation (\ref{eq:JJ})
states clearly that there are two different contribution to, in fact,
any quantity, namely the {\em valence} part coming from the explicit
quark propagator ($E_{\rm level}$), and the {\em sea} contribution
corresponding to the effective action ($E_{\rm field}$). 

 Let us rewrite the effective action
(\ref{eq:Seff}) in terms of the Euclidean spectral representation
in a form ready for expansion in $\Omega$ and $m$:
\bea
S_{\rm eff} & = &  -N_{\rm c}T \, \int\frac{d\omega}{2\pi}
{\rm Tr} \log\, ( i \omega+H )
\left[ 1+\frac{1}{i\omega+H}(-i\gamma_{4} A^{\dagger} m\,A+A^{\dagger}\dot{A})
\right] \label{eq:Slog}
\end{eqnarray}
where H is the hermitian static hamiltonian:
$ H=-i\gamma_{4} (-i\gamma_{i} \partial_{i}+M U_{0})$.
The first term corresponds to the static soliton energy. Integrating by
parts and subtracting the vacuum contribution we get:
\beq
S_{\rm eff} =  N_{\rm c} T \, \int \frac{d \omega}{2 \pi}
{\rm Tr} \left(\frac{i \omega}{i \omega + H}-\frac{i \omega}{i \omega +H_{0}} 
\right)  
  =  \frac{N_{\rm c} T}{2} \sum\limits_{n}  
 (\mid E_{n} \mid - \mid E_{n}^{0} \mid ). \label{eq:Msol}
\eeq
So the static energy of the soliton is given by a sum of the
differences of energies of the levels of the static hamiltonian H with
and without the soliton. This sum has to be of course regularized and
the contribution of the valence level has to be added. 
 Once the static energy of
the soliton is found, selfconsitently or by variational methods, we can
proceed further and expand  Eq.(\ref{eq:Slog}) in
powers of $\Omega$ and $m$. 

Not all terms in this expansion should be regularized.
Those which come from the {\em imaginary } part of
the effective action need not any  regularization. 
The imaginary part is connected to the anomalous
term, or in other words, to the Wess-Zumino term. As an example
let us consider a term linear in 
$ \Omega_{a}$:
\begin{eqnarray}
S_{\rm eff} &= & -N_{\rm c}T \, \int\frac{d\omega}{2\pi}
{\rm Tr} 
\left( \frac{1}{i \omega+H} \lambda_{a} - 
       \frac{1}{i \omega+H_{0}} \lambda_{a} \right) \Omega_{a}
\nonumber \\
&= & - i N_{\rm c}T \,\frac{1}{2 \sqrt{3}} \; \Omega_{8}\,
\frac{1}{2} {\rm Tr}\left( {\rm sign}(H) - {\rm sign}(H_{0}) \right) 
\label{eq:WZ}
\end{eqnarray}
That only $\Omega_8$ contributes is due to the fact that $H$
is just the $SU(2)$ operator and what survies the trace is $\lambda_8$.
The quantity
$ 1/2\, {\rm Tr}\left( {\rm sign}(H) - {\rm sign}(H_{0}) \right) $ 
counts the  number of levels that crossed $E=0$ line.  The
profound connection of Eq.(\ref{eq:WZ})  to the topology of our
{\em hedgehog} Ansatz is probably clear to all readers familiar with the
Skyrme model$^{11),12)}$.\nc{skm:W1,skm:W2}

\section{SEMICLASSICAL QUANTIZATION}

Let us expand $S_{\rm eff}$ up to the quadratic order in 
$\Omega$ (in Minkowski metric):
\begin{eqnarray}
L_0 &=&-M_{\rm cl}[P]+\frac{I_{A}[P]}{2} \sum_{i=1}^{3} \Omega_{i}^{2}
  +  \frac{I_{B}[P]}{2} \sum_{k=4}^{7} \Omega_{k}^{2}
  - \frac{N_{\rm c}}{2 \sqrt{3}} \Omega_{8} \label{eq:L} 
\end{eqnarray}
where:
\begin{eqnarray}
I_{ab}= - \frac{N_{\rm c}}{4}\int\frac{d\omega}{2\pi} {\rm Tr}
\left[
\frac{1}{i\omega+H}\lambda_{a}\frac{1}{i\omega+H}\lambda_{b}
\right] = \left\{
\begin{array}{ccl}
I_{A}\delta_{ab} & {\rm for} & a,b=1...3 \\
I_{B}\delta_{ab} & {\rm for} & a,b=4...7~. \\
       0         & {\rm for} & a,b=8 
\end{array}\right. \label{eq:IAB}
\end{eqnarray}
This lagrangian reminds the Skyrmion lagrangian. The quantization
proceeds as in the Skyrme model case and the hamiltonian reads:
\beq
H_0 = M_{\rm cl} + H_{\rm SU(2)} +H_{\rm SU(3)} ,
\eeq
\begin{eqnarray}
H_{\rm SU(2)}=\frac{ 
C_{2}( {\rm SU(2) }
} {2I_{A}}, & &  
H_{\rm SU(3)}=
\frac{ 
C_{2}({\rm SU(3)})-C_{2}({\rm SU(2)})-\frac{N_{\rm c}^{2}}{12}  
}
{2I_{B}}.   \nonumber
\end{eqnarray}
Here SU(3) is the flavor and SU(2) the rotational symmetry.
It can be shown that the baryon wave functions are given in terms of
SU(3) Wigner D functions:
\begin{eqnarray}
\psi_{\rm baryon}^{(\mu)}(A) 
 & = &\sqrt{{\rm dim}(\mu)}
\left< Y,I,I_{3} \mid D^{(\mu)}(A) \mid Y_{\rm R} ,J,-J_{3}
\right>^{\bf *},
\label{eq:D}
\end{eqnarray}
where $Y_{\rm R}$ is in fact constrained due to the linear term in
$\Omega_8$ to be 1. The lowest SU(3) representations which contain
states with $Y=1$ are$^{13)}$:\nc{skm:G} $\mu=${\bf 8} and $\mu=${\bf 10}. 
 
In order to fix {\bf 10} -- {\bf 8} splitting we need 
the constituent quark mass 
$M\approx 395$~MeV, certainly a reasonable number$^{1})$. What comes out too
high is the absolute mass of the soliton $M_{\rm cl}=1.2$~GeV.
 There are however some negative corrections to it, which are
of the order $O(1)$ whereas $M_{\rm cl}$ is $O(N_{\rm c})$. Instead on
insisting on the calculation of the absolute masses let us
concentrate on the mass splittings. 

\section{MASS SPLITTINGS}

To calculate mass splittings one has to expand Eq.(\ref{eq:Slog}) 
 in powers of the current quark mass $m=
\mu_{0}\,\lambda_{0} - \mu_{8}\,\lambda_{8} -\mu_{3}\,\lambda_{3}
$ ($\lambda_{0}=\sqrt{2/3} \;{\bf 1}$)  where:
\beq
 \mu_{0}  =  \frac{1}{\sqrt{6}}(m_{\rm u}+m_{\rm d}+m_{\rm s}),
~~\mu_{8}  = \frac{1}{\sqrt{12}}(2\,m_{\rm s}-m_{\rm u}-m_{\rm d}), 
~~\mu_{3}  =  \frac{1}{2}(m_{\rm d}-m_{\rm u}).
\eeq

There
will be terms of the order of $m$, $m^2$ and mixed terms: $m \Omega$.
Note that:
$ A^{\dagger}\lambda_{a}\;A  =  D^{(8)}_{a b}(A)\;
 \lambda_{b} 
$.
So we get lagrangian (in Minkowski space):
\begin{eqnarray}
L_{m}~ & = &-\sigma m_{\rm s} + \sigma ( m_{\rm s} D_{88}^{(8)}
        + \frac{\sqrt{3}}{2} \Delta m D_{38}^{(8)}),  \nonumber\\
 L_{m \Omega} & = & -\frac{2}{\sqrt{3}} m_{\rm s} D_{8a}^{(8)}
K_{ab}{\Omega}_{b}  -\Delta m D_{3a}^{(8)} K_{ab}{\Omega}_{b},  \nonumber \\
L_{m^2} & = & \frac{2}{9} m_{\rm s}^2 
(N_{0} (1-D_{88}^{(8)})^2+ 3 N_{ab} D_{8a}^{(8)} D_{8b}^{(8)})
\nonumber \\
& & +\frac{2}{3\sqrt{3}} m_{\rm s}\Delta m 
(-N_{0} (1-D_{88}^{(8)}) D_{38}^{(8)}
+ 3 N_{ab} D_{3a}^{(8)} D_{8b}^{(8)} )
\end{eqnarray}
where constant $\sigma$
is related to the  sigma term 
$\Sigma=3/2(m_{\rm u}+m_{\rm d})\sigma$ (
$\Sigma=58$~MeV in this model). 
Similarly to tensor $I_{ab}$ of Eq.(\ref{eq:IAB}) $\sigma$, $N_{ab}$ and
{\em anomalous} 
tensor  $K_{ab}$ 
are given in terms of traces over certain Dirac
and/or flavor matrices and denominators $1/(i\omega+H)$. Their explicit
forms will be given elswhere$^{14)}$.\nc{mp:ms2}
\begin{figure}
\vspace{-2cm}
\centerline{\psfig{figure=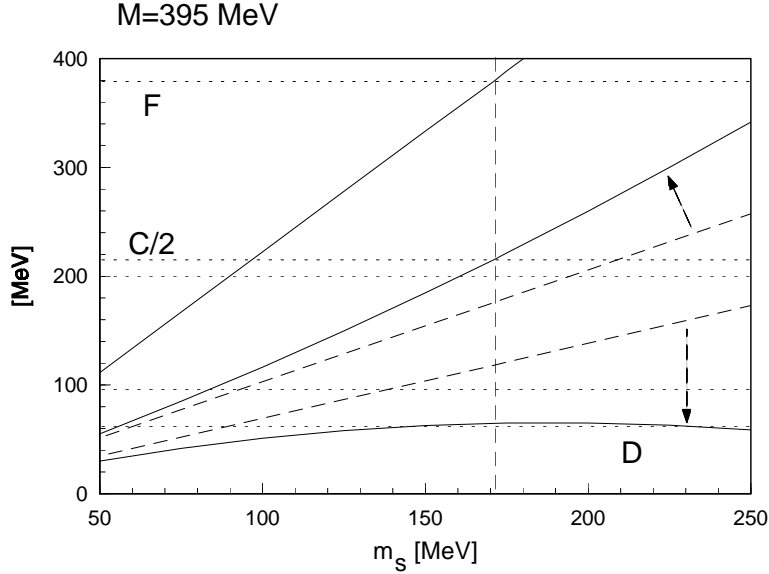,height=16 cm}}
\vspace{-7cm}
\caption{Parameters $F$, $D$ and $C$ as functions of the strange quark
 mass} 
 \end{figure} 

It is a matter of simple algebra to derive the pertinent hamiltonian
and calculate the hyperon 
mass splittings up to the second order in $m_{\rm s}$.
Let us note that up to terms linear in current quark masses 
the Gell-Mann Okubo
(GMO)
mass formulae are reproduced:
\begin{eqnarray}
\Delta M_{B}^{(8)}  =  -\frac{F}{2}{\rm\bf Y} -
                          \frac{D}{\sqrt{5}} 
                          ( 1 - {\rm\bf I}^{2} +
                          \frac{1}{4}{\rm\bf Y}^{2}),
                    &  &                                                    
\Delta M_{B}^{(10)}  =  -\frac{C}{2 \sqrt{2}}{\rm\bf Y}. 
\end{eqnarray}
In the order $O(m_{\rm s}^2)$ terms breaking 
GMO parametrization appear. They are however numerically negligible. In
Fig.1 we plot GMO constants $F$,$D$ 
and $C$. Horizontal dashed lines represent
experimentally allowed ranges. Long-dashed lines correspond to terms
linear in $m_{\rm s}$, whereas solid lines include $O(m_{\rm s}^2)$
corrections. Since the $O(m_{\rm s}^2)$ 
correction to $F$ is negligible, and since $F$ is
given in terms of only one mass difference ($\Xi-$N) so that 
there are no experimental
errors to it, we choose $m_{\rm s}=171$~MeV (vertical dashed line) to
reproduce its experimental value. The arrows indicate the influence of
the quadratic mass corrections on $D$ and $C$; they substantially
improve agreement with experiment.

In Fig. 2. we plot the hadronic parts of isospin splittings for the
octet$^{2),14)}$.
The long-dashed lines represent contribution linear in $\Delta m =
m_{\rm d}-m_{\rm u}$, while solid lines include $m_{\rm s}$
correction. In the order $O(m_{\rm s}^0\Delta m)$ 
the experimental ranges for
the isospin splittings are reproduced by $\Delta m\approx 3.5$~MeV (two
vertical lines). The arrows indicate shift due to the 
$O(m_{\rm s}\Delta m)$
terms. Here the common range shrinks to a value in a vicinity of
$\Delta m\approx 4.3$~MeV. Taking into account the simplicity of the
model this result seems to be surprisingly good.
\begin{figure}
\vspace{-2cm}
\centerline{\psfig{figure=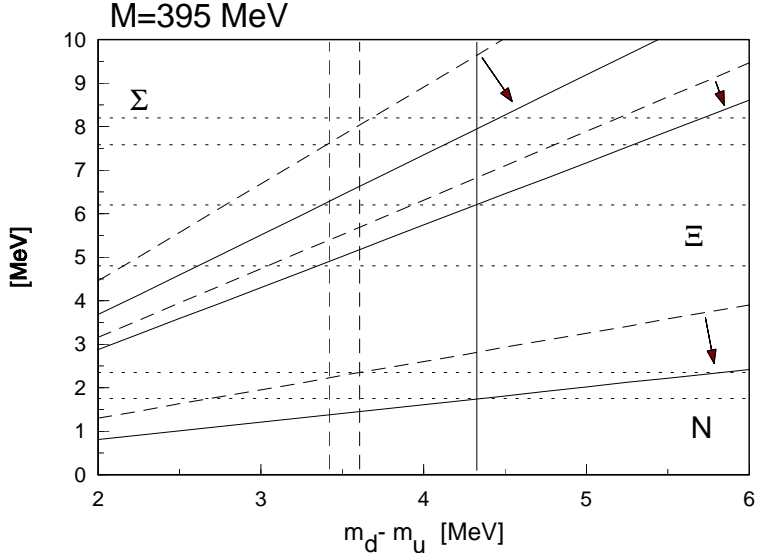,height=16 cm}}
\vspace{-7cm}
\caption{Hadronic parts of isospin splittings as functions of the 
down-up mass difference for different values of the strange quark mass}
 \end{figure}

\section{FINAL REMARK: AXIAL COUPLINGS }

We have developed a systematic approach to calculate
nonperturbative quantities in the semibosonized NJL (or chiral quark) model.
Since the mass splittings in the simplest version of the model (only
one scalar coupling, no vector mesons, no gluonic corrections, no boson
loops) show surprisingly good agreement with  experiment it is tempting
to apply it to other observables. As an example let us consider the
axial form-factors of hyperons. Let us first note that the very
recent result on $g_{\rm A}^{(3)}$ seems to solve the long lasting 
problem
of underestimation of this quantity in NJL model$^{15)}$.
\nc{ga:WaWa} Similarly to the mass
splittings $g_{\rm A}^{(3)}$ is a series in $\Omega$ and $m$. For a long
time it has been overlooked that there is a non-zero contribution to
$g_{\rm A}^{(3)}$ linear in $\Omega$. This contribution shifts the
zeroth order value of approximately 0.8 by 0.4. Although this
correction is large it acts in a right direction and puts the
previously obtained result for a singlet axial current$^{16)}$ \nc{ga:BPG}
 $g_{\rm A}^{(0)}\approx 0.4$ 
(to which there are no $O(\Omega)$ corrections) on much
safer ground. The above result should be compared with the EMC result
of $0.13 \pm 19\;^{17)}$.\nc{ga:JM} One should not forget though, that 
the experiment was done
at the renormalization scale $<Q^2>=10$~GeV$^2$. 
The evolution backwards in $Q^2$,
although  weak, tends to increase this value. So it is clear
that both numbers are consistent with each other. The contribution 
to $g_{\rm A}^{(0)}$ comes almost entirely from the valence part. This
is consitent with the Skyrme model zero result for this quantity$^{18)}$. 
\nc{ga:BEK} On
the other hand the model clearly shows that
the matrix element of the singlet axial current is
substantially less than 1 in contrast to the naive quark model. 


\baselineskip=12pt
\begin{thebibliography}{10}
\bibitem{cqm:Blotz2}
A.~Blotz, D.~I. Diakonov, K.~Goeke, N.~W. Park, V.~Petrov, and P.~V. Pobylitsa,
\newblock The {S}{U}(3)-{N}ambu--{J}ona-{L}asinio soliton in the collective
  quantization formulation,
\newblock Bochum Univ. preprint RUB-TPII-27/92, 1992,
\newblock Nucl. Phys. A, in print.
\vb \bibitem{mp:iso}
M.~Prasza{\l}owicz, A.~Blotz, and K.~Goeke,
\newblock {\em Phys. Rev.}, {\bf D47} (1992) 1127.
\vb \bibitem{chl:Ball3}
R.~Ball,
\newblock {\em Phys. Rep.}, {\bf 182} (1989) 1.
\vb \bibitem{chl:IanMog}
I.~J.~R. Aitchison,
\newblock Effective lagrangians for low energy hadron physics,
\newblock In M.~Je{\.z}abek and M.~Prasza{\l}owicz, editors, {\em Workshop on
  {S}kyrmions and Anomalies}, page~5; World Scientific, 1987.
\vb \bibitem{mp:padova}
M.~Prasza{\l}owicz,
\newblock {\em Il Nouv. Cim.}, {\bf A102} (1989) 39.
\vb \bibitem{inst:dp1}
D.~I. Dyakonov and V.~Yu. Petrov,
\newblock {\em Nucl. Phys.}, {\bf B245} (1984) 259.
\vb \bibitem{inst:dp2}
D.~I. Dyakonov and V.~Yu. Petrov,
\newblock {\em Nucl. Phys.}, {\bf B272} (1986) 457.
\vb \bibitem{mp:ga}
P.~Sieber, M.~Prasza{\l}owicz, and K.~Goeke,
\newblock Dependence of baryonic observables on the quark axial-vector coupling
  in chiral quark model,
\newblock Bochum Univ. preprint RUB-TP2-38/93, 1993.
\vb \bibitem{rvs:bal}
A.~P. Balachandran,
\newblock Syracuse University preprint, 1987,
\newblock SU-4428-361.
\vb \bibitem{inst:nucleon}
D.~I. Dyakonov, V.~Yu. Petrov, and P.~V. Pobylitsa,
\newblock {\em Nucl. Phys.}, {\bf B306} (1988) 809.
\vb
\vb \bibitem{skm:W1}
E.~Witten,
\newblock {\em Nucl. Phys.}, {\bf B223} (1983) 422.
\vb \bibitem{skm:W2}
E.~Witten,
\newblock {\em Nucl. Phys.}, {\bf B223} (1983) 433.
\vb \bibitem{skm:G}
E.~Guadagnini,
\newblock {\em Nucl. Phys.}, {\bf B236} (1984) 35.
\vb \bibitem{mp:ms2}
A.~Blotz, M.~Prasza{\l}owicz, and K.~Goeke,
\newblock in preparation.
\vb \bibitem{ga:WaWa}
M.~Wakamatsu and T.~Watabe,
\newblock The $g_{\rm {A}}$ problem in the hedgehog soliton models and its
  resolution,
\newblock Osaka Univ. preprint, February 1992.
\vb \bibitem{ga:BPG}
A.~Blotz, M.V. Polyakov, and K.~Goeke,
\newblock {\em Phys. Lett.}, {\bf B302} (1993) 151.
\vb \bibitem{ga:JM}
R.L. Jaffe and A.V. Manohar,
\newblock {\em Nucl. Phys.}, {\bf B337} (1990) 509.
\vb \bibitem{ga:BEK}
S.~Brodsky, J.~Ellis, and M.~Karliner,
\newblock {\em Phys. Lett.}, {\bf 206B} (1988) 309.
\end{thebibliography}

\end{document}